\documentclass[runningheads]{llncs}
\usepackage{geometry}
\usepackage[T1]{fontenc}
\usepackage{graphicx}
\usepackage[utf8]{inputenc}
\usepackage{comment}
\usepackage{soul, color}
\usepackage{algorithm2e}
\usepackage{listings}
\usepackage{multirow}
\usepackage{caption}
\usepackage{subcaption}

\newcommand{\tweets}{5,583,168 }
\newcommand{\users}{1,858,605 }

\begin{document}

\title{Characterizing the 2022 Russo-Ukrainian Conflict Through the Lenses of Aspect-Based Sentiment Analysis: Dataset, Methodology, and Preliminary Findings.}

\author{Maurantonio Caprolu\inst{1} \and
Alireza Sadighian\inst{1} \and
Roberto Di Pietro\inst{1}}
\authorrunning{M. Caprolu et al.}
\titlerunning{Characterizing the 2022 Russo-Ukrainian Conflict on Twitter}
%
\institute{Division of Information and Computing Technology, College of Science and Engineering, Hamad Bin Khalifa University, Qatar Foundation, Doha, Qatar}


\maketitle

\begin{abstract}
Online social networks (OSNs) play a crucial role in today's world. On the one hand, they allow free speech,  information sharing, and social-movements organization, to cite a few. 
On the other hand, they are the tool of choice to spread disinformation, hate speech, and to support propaganda. For these reasons, OSNs data mining and analysis aimed at detecting disinformation campaigns that may arm the society and, more in general, poison the democratic posture of states, are essential activities during key events such as elections, pandemics, and conflicts.

In this paper, we studied the 2022 Russo-Ukrainian conflict on Twitter, one of the most used OSNs. We quantitatively and qualitatively analyze a dataset of more than 5.5+ million tweets related to the subject, generated by 1.8+ million unique users. By leveraging statistical analysis techniques and aspect-based sentiment analysis (ABSA), we discover hidden insights in the collected data and abnormal patterns in the users' sentiment that in some cases confirm while in other cases  disprove common beliefs on the conflict. In particular, based on our findings and contrary to what suggested in some mainstream media, there is no evidence of massive disinformation campaigns.
However, we have identified several anomalies in the behavior of particular accounts and in the sentiment trend for some subjects that represent a starting point for further analysis in the field. 
The adopted techniques, the availability of the data, the replicability of the experiments, and the preliminary findings, other than being interesting on their own, also pave the way to further research in the domain.

\keywords{Online Social Networks  \and Aspect-Based Sentiment Analysis \and Twitter \and Data Mining \and Data Analysis \and 2022 Russo-Ukrainian Conflict}   
\end{abstract}

\section{Introduction}
\label{sec:introduction}


Online Social Networks (OSN) were born   to connect  users so that they could share their emotions, desires, happiness, hobbies and interests as an important aspect of their socialization. While this is still true, OSNs have also evolved towards platforms were political opinions are formed and discussed, and where groups organize themselves in order to achieve shared objectives.
The most successful  platforms are characterized by being weakly moderated, and  designed to publish content 
much faster than any classic social media (newspapers, magazines, etc.). 
Hence, on the one hand OSNs play a key role in forming the public opinion about almost any event. 
On the other hand,  they are the ideal platform for spreading fake news and disinformation~\cite{ndiw2021}.

In this paper, we conduct a detailed study on the 2022 Russo-Ukrainian conflict. Our main goal is to characterize the sentiment related to the escalation of the conflict for a specific OSN: Twitter. 
Moreover, we also searched for digital clues that could suggest possible disinformation campaigns and mined the collected data to highlight possible anomalies in the analyzed accounts. To reach these objectives, we built  a massive dataset on the subject by collecting \tweets tweets shared by \users different users between January 27 and March 23, that is, one month before and one month after February 24, 2022. On that day, the Russian president Vladimir Putin announced the start of a special military operation in Ukraine. Our working hypothesis is that if there was a disinformation campaign aimed at manipulating the public opinion of English-speaking countries on the conflict in Ukraine, as suggested by several mainstream media\footnote{https://www.theguardian.com/world/2022/apr/07/propaganda-social-media-surge-invasion-ukraine-meta-reports}, its traces should be captured by our dataset.
Therefore, we performed a preliminary analysis to discover any hidden abnormal pattern that might suggest such a disinformation campaign. In particular, we applied quantitative and qualitative analysis on the collected data using statistical analysis techniques and an advanced Natural Language processing (NLP) technique, called Aspect-Based Sentiment Analysis (ABSA). ABSA allows to extract aspects (the feature, entity, or topic that is being talked about) and to associate a sentiment to them  (Positive, Negative, Neutral)~\cite{nazir2020issues}. 
The analysis over time of these data unleashed a trove of insights on the public perception of the conflict and its actors.
Using the cited techniques, we also studied the public perception of the Russo-Ukrainian conflict on twitter. In particular, we compute the evolution over time of the sentiment for  5 main keywords: ``Putin'', ``Zelensky'', ``NATO'', ``Ukraine'' and ``Russia'', respectively, identifying them  as the main ABSA aspect terms expected to provide  
meaningful insights for the conflict at study.
The knowledge gained by our analysis  provides some confirmations as well as falsifications of current believes on the conflict. In particular, as per confirmations, it appears that the negative sentiment related to President Putin increased after the escalation of the conflict, the neutral sentiment towards him decreased by roughly the same amount, while the positive one remained stable.
A specific insight is that this negative sentiment does not appear pegged to the whole concept of Russia, signaling the fact that users are somehow able to distinguish between the action of a government and the subjected people. 
A  common belief is the fact that the Russia's presence on social media, and especially troll and bot, is strong 
\cite{kiriya2021troll}.
However, from our data, it does not appear that such activity took place, in the Twitter platform, for the studied conflict.  

{\bf Contributions.} The major contributions of this research are as follows: 
\begin{itemize}
    \item We built a dataset for the 2022 Russo-Ukrainian conflict by collecting more than 5.5 million tweets related to the subject. Although we cannot publish it due to Twitter's policies, we reported the methodology and the query we used for this purpose in Section~\ref{sub:data_collection}, allowing any Twitter Developer Account to easily download and reconstruct exactly the same dataset.
    \item We classified the user accounts into five different categories and analyzed those accounts  according to several metrics.
    \item We performed a preliminary analysis of our dataset, resorting to ABSA, to characterize the sentiment about the conflict shared on Twitter in the English-speaking world.
    \item We discussed our results revealing statistics and sentiment trends for the major players involved in the conflict and provided valuable insights for future investigation on disinformation related to the Russo-Ukrainian conflict.
    \item The replicability of the experiments and the novel techniques adopted, joined with the gained insights and the highlighted future work, pave the way for further research on the comparative strength and weaknesses among ABSA and other data mining techniques.
\end{itemize}

\textbf{Roadmap.} The rest of this paper is organized as follows. In Section~\ref{sec:related}, we discuss the related work. In Section~\ref{sec:methodology}, we present our proposed approach in detail. We show our results in Section~\ref{sec:results}, and we conclude in Section~\ref{sec:conclusion} with some insights for future research.

\section{Related Work}
\label{sec:related}
The 2022 Russo-Ukrainian conflict 
is a very recent event at the time of writing. Nevertheless, several studies have already leveraged the OSNs to gain insight on some aspects of the conflict, demonstrating the need to identify a dataset for supporting different research directions on this topic. 

Although existing work shared the same goal, i.e., to build a dataset of OSN contents on the Russo-Ukrainian conflict, methodologies and results are pretty different. In~\cite{haq2022twitter}, the authors provide a collection of tweets (in English language only) built by using a set of keywords that changes over time according to the main events of the conflict. A similar work is reported in~\cite{pohl2022twitter}, with static keywords and a small post-processing phase to discard tweets non-related to the subject. A collection of raw data extracted from Twitter was proposed also in~\cite{chen_ferrara}, where authors conducted a small analysis on data volume. In~\cite{park2022voynaslov}, instead, the authors pursued the same goal, but targeting information shared in Russian language, retrieved from both a Russian state-affiliated platform (VK) and Twitter. To the best of our knowledge, we are the first to build a dataset with more refined data, e.g., original contents only, and extensive post-processing, e.g., user categorization.

Some pioneering works leveraged ABSA for analyzing OSN data on several domains, such as education~\cite{sivakumar2017aspect}, medicine~\cite{bhatia2020aws}, and marketing~\cite{gamzu2021identifying}. However, to the best of our knowledge, we are the first to apply ABSA to the Russo-Ukrainian conflict and, more in general, to detect abnormal patterns in sentiment trends that may suggest disinformation activities.




\section{Methodology}
\label{sec:methodology}

To characterize the 2022 Russo-Ukrainian Conflict on Twitter, we design a methodology to: 1) build the appropriate dataset, 2) analyze both quantitatively and qualitatively all the tweets collected on this topic.
\vspace{-0.5cm}
\begin{figure}
\centering
\includegraphics[width=0.8\columnwidth]{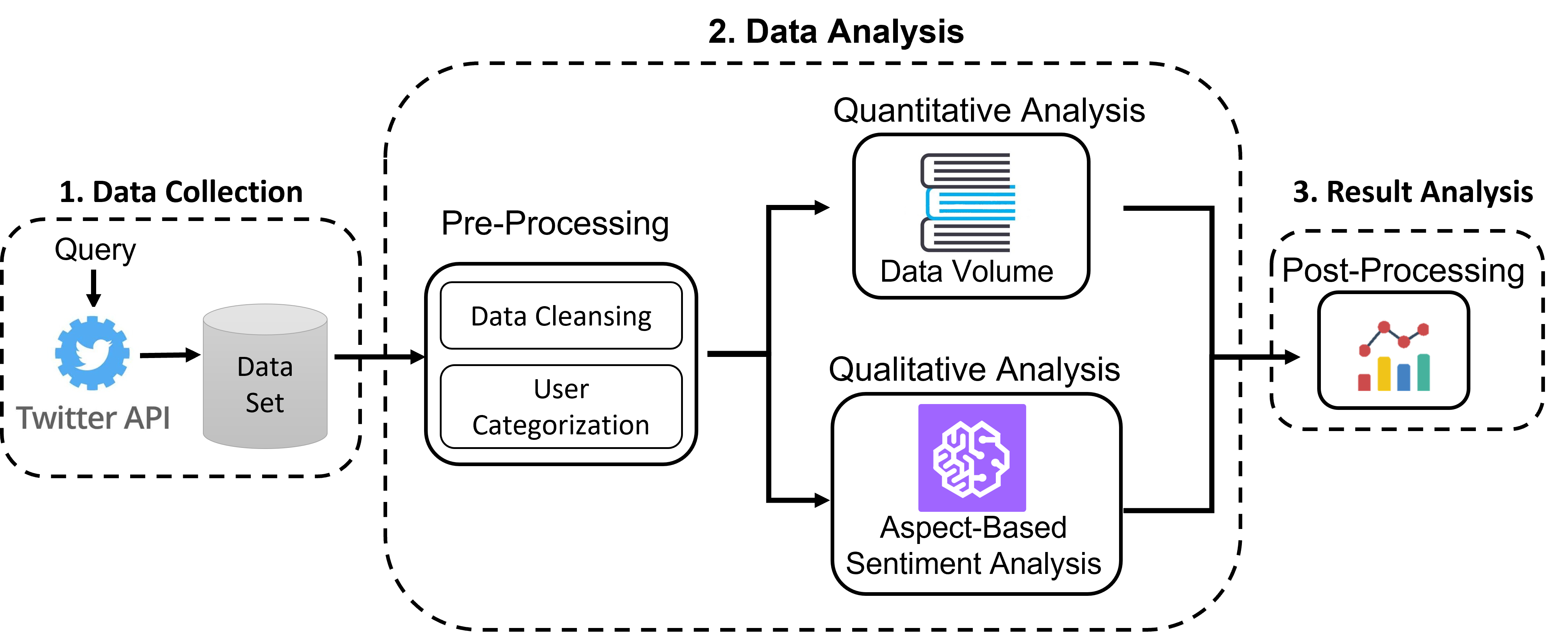}
\caption{Research Methodology} 
\label{fig:methodology}
\vspace{-0.5cm}
\end{figure}

\noindent

Figure~\ref{fig:methodology} illustrates in detail our approach consisting of three different phases. The first step is data collection. To build our dataset, we collected tweets related to the target topic via the Twitter API. Next, after pre-processing our data as detailed in Sec.~\ref{sub:pre_processing}, we analyze the collected tweets from both the quantitative and qualitative perspectives. In particular, we evaluated the volume of our data, and we investigated the user's sentiment over time by leveraging Aspect-Based Sentiment Analysis, as described in Sec.~\ref{sub:sentiment}. Each analyzed dimension produces different insights and results, that we analyzed in Sec.~\ref{sec:results}, while also providing  preliminary findings. 

\subsection{Data Collection} \label{sub:data_collection}
To build our dataset, we collected tweets related to the Russo-Ukrainian Conflict via the Twitter API v2. Specifically, we queried the ``tweets/search/all'' Twitter endpoint\footnote{https://api.twitter.com/2/tweets/search/all}, 
which returns the complete history of public tweets matching a search query, from the very first tweet in the platform (March 26, 2006) to the last one (query execution time)\footnote{https://developer.twitter.com/en/docs/twitter-api/tweets/search/api-reference/get-tweets-search-all}.

To select only the tweets that would best highlight changes in the public sentiment,  
first, we have narrowed the search space to two months, one month before and one month after February 24, 2022. On that day, the Russian president, Vladimir Putin, announced the start of a special military operation in Ukraine. Then, we have selected several keywords, reported in Listing~\ref{lst:keywords}, to define our area of interest.

\begin{lstlisting}[caption={The selected keywords   for data collection.}, captionpos=b,label={lst:keywords},language=SQL, 
           showspaces=false,
           basicstyle=\footnotesize\ttfamily,
           frame=bt,
           commentstyle=\color{gray}]
zelensky, ukraine, ukrainian, russia, russian, putin
\end{lstlisting}

As we explore the topic of the Russo-Ukrainian conflict, we are also interested in the perception of the reference users over the North Atlantic Treaty Organization (NATO). However, we did not use that keyword for data collection because it is not directly involved in the topic under study. In fact, by including the NATO keyword, we would also have collected tweets that do not refer, neither directly nor indirectly, to the Russo-Ukrainian Conflict. Nevertheless, we considered the NATO keyword in the data analysis by verifying both the volume and sentiment of the collected tweets that named the Atlantic alliance along with one of the keywords in Listing~\ref{lst:keywords}.

Finally, we used some operators to refine further the results returned by the Twitter endpoint. In particular, we decided to target original content only, avoiding retweets, replies, and quotes. 
Moreover, as we are only interested in textual content, we excluded tweets containing links and media in the tweet body. Also, considering the tools and the scope of this research, we analyzed only tweets in English.\\

\begin{lstlisting}[caption={The query we sent via GET request to the Twitter API v2, ``tweets/search/all'' endpoint.}, captionpos=b,label={lst:query},language=SQL, 
           showspaces=false,
           basicstyle=\footnotesize\ttfamily,
           frame=bt,
           numbers=left,
           numberstyle=\tiny,
           commentstyle=\color{gray}]
start_time = "2022-01-27T00:00:00.000Z"
end_time = "2022-03-23T00:00:00.000Z"

(zelensky OR zelenskyy OR ukraine OR ukrainian 
OR russia OR russian OR putin) 
-has:links -has:media -is:retweet -is:reply -is:quote lang:en
\end{lstlisting}

\noindent
The query we used to build our dataset is reported in Listing~\ref{lst:query}. We inserted two parameters in the GET request to restrict the time space of the search from January 27 to March 23, i.e., 1 month before and 1 month after the triggering of the conflict  (lines 1 and 2). Our keywords are in OR logic to catch all tweets that contain at least one of them (lines 4 and 5). We excluded retweets, replies, and quotes by using the ``is'' operator prepended by a dash to negate it (line 6). Similarly, we used the (negated) ``has'' operator to exclude tweets that include any link or share any media (line 6). 
By executing the query described above, we collected \tweets tweets from \users different users.



\subsection{Pre-processing}
\label{sub:pre_processing}
After building our dataset, we first cleaned up the data returned by the Twitter API to remove tweets that fall outside the scope of this research, or contain anything other than English text that could undermine subsequent analyses. We removed a total of 1144 tweets, 856 of which, although returned by the Twitter API, did not actually contain any of the keywords used in the query---this latter behaviour was duly noted down for further analysis in future works. 
Then, we categorized Twitter accounts based on different metrics to understand the user profiles that are more interested on the topic. In particular, we divided all the users included in our dataset into the following five categories:
\begin{itemize}
    \item \textbf{Trusted Accounts:} This category includes all the accounts for which the owner's real identity is, somehow, publicly known. We put in this category accounts flagged as ``verified'' by Twitter 
    and/or accounts with a high number of followers. Specifically, we considered users with celebrity status, i.e. $>9M$ followers, and very popular account, i.e. $900K$ to $1.1M$ followers~\cite{10.1145/3110025.3110090}.
    \item \textbf{Baby Accounts:} This category includes all accounts that can be considered ``young'' during this study, i.e. account generated from September 2021 onward. We consider this category interesting as it may include a subset of accounts explicitly created for tweeting about the escalation in the Russo-Ukrainian conflict. 
    \item \textbf{Abnormal Accounts:} This category contains accounts that do not behave similar to regular users. Such accounts are suspected to be managed by bots, trolls, and possibly other unconventional users. To identify this type of account, we used a standard metric in the literature \cite{chu2010tweeting}, the Friend Ratio (FR), which leverages two of the few user metrics provided by Twitter---the number of followers and the number of following. In particular, for each account, we compute the Friend Ratio (FR) as the ratio of follower/following. Then, we considered as ``abnormal'' three different situations that we believe are unlikely to occur in accounts managed by private citizens : (i) accounts with zero following; (ii) accounts with an extremely low FR, i.e., less than or equal to 0.02, as considered bots with high probability by existing works\footnote{https://i.blackhat.com/us-18/Wed-August-8/us-18-Anise-Wright-Dont-@-Me-Hunting-Twitter-Bots-at-Scale-wp.pdf}; and, (iii) accounts with a fair FR, i.e., $>= 0.99$ and $<= 1.1$, as normal users tend to have a higher number of followers with respect to their following, since they follow celebrities or other popular accounts that don't follow them back. Users who fall into this category but were also created after September 1, 2021, have been included in the baby accounts category.
    \item \textbf{Unknown Accounts:}  This category includes accounts for which Twitter did not return information because the user was suspended or deleted from the platform at the time of the data collection.
    \item \textbf{Regular Accounts:} This category includes all remaining accounts that do not fall into the categories discussed above.
\end{itemize}

\subsection{Aspect-Based Sentiment Analysis}
\label{sub:sentiment}

\begin{figure}
\centering
\includegraphics[width=0.7\columnwidth]{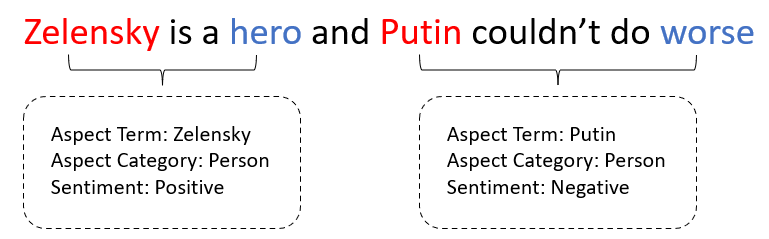}
\caption{An example of ABSA analysis}
\label{fig:ABSA_example}
\vspace{-0.5cm}
\end{figure}

Aspect-based sentiment analysis (ABSA) helps identifying fine-grained opinion polarity toward a specific aspect associated with a given target. Unlike general sentiment analysis, ABSA can provide more detailed information helping to accomplish more advanced analysis~\cite{xu2019bert}. 
Figure~\ref{fig:ABSA_example} illustrates how ABSA works for a sample sentence. 
In this example, the result of general sentiment analysis is \textit{``Mixed''} as there is one negative sentiment and one positive sentiment. 
\vspace{-0.5cm}
However, as the figure illustrates, the sentiment for the ABSA aspect term ``Zelensky'' is positive while it is  negative for ``Putin''. 
Hence, general sentiment analysis would allow us to identify only the prevalent sentiment of a single tweet. ABSA, instead, allows us to extract the different sentiments related to subjects identified in a single tweet, hence enabling a more comprehensive, fine grained analysis.
Since our goal is to understand the sentiment related to different subjects involved over the cited period (starting from one month before the conflict, till one month after), we have chosen to use ABSA to get more information from every single tweet.

\textbf{Tools.} 
Considering that ABSA is still an ongoing research topic in NLP field, there are not many open-source or third-party solutions providing this service in industry or academia. We considered two major solutions for this study: (i) A Python library called aspect-based-sentiment-analysis 2.0.3\footnote{https://pypi.org/project/aspect-based-sentiment-analysis/}
and (ii) Amazon AWS Comprehend Targeted Sentiment Analysis\footnote{https://docs.aws.amazon.com/comprehend/latest/dg/how-targeted-sentiment.html}. 
{\em AWS Comprehend} provides more detailed results, such as sentiment scores, aspect term type, aspect term location, etc. whereas, the Python library provides only aspect terms and corresponding sentiments. Hence, we chose to conduct our experiments with {\em AWS Comprehend} which has been widely used by researchers~\cite{gamzu2021identifying}\cite{bhatia2020aws}\cite{bhatia2019comprehend}.


\subsection{Post-Processing and Results Generating}
\label{sub:post_processing}
\begin{figure}
\centering
\includegraphics[width=0.8\columnwidth]{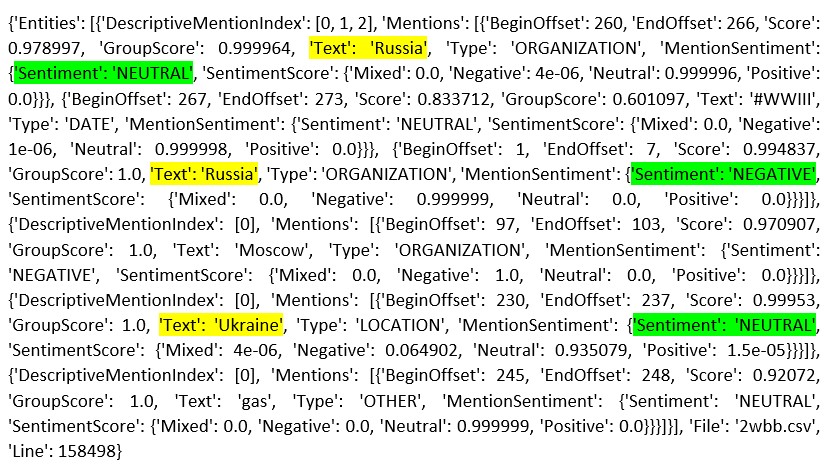}
\caption{ABSA raw result showing aspect terms and corresponding sentiments}
\label{fig:raw_result}
\vspace{-0.5cm}
\end{figure}

 The input dataset to the {\em AWS Comprehend} is a CSV file including the list of tweet texts to be analyzed. Considering the {\em AWS Comprehend} input limits of 30,000 lines in each input file, we split our dataset into a group of CSV files and submit them to {\em AWS Comprehend} in an ordered sequence. The {\em AWS Comprehend} outputs are JSON files. For each tweet, {\em AWS Comprehend} provides a raw JSON entity including several aspect terms (subjects, objects, nouns, etc.) of that tweet, their corresponding sentiments and the sentiment scores.  Figure~\ref{fig:raw_result} illustrates a screen shot of {\em AWS Comprehend} output result. The yellow highlighted terms are the identified aspect terms, and the green highlighted terms are the corresponding sentiments.

In this study, we selected the five keywords reported in Listing~\ref{lst:keywords_ABSA} as the major aspect terms involved in the topic of our interest.
Then, we used {\em AWS Comprehend ABSA} to extract the sentiment related to them from all the tweets in our dataset.
For this purpose, we first submitted our dataset (\tweets tweets in CSV file format) to the {\em AWS Comprehend} platform as input. Next, we received corresponding ABSA results (aspect terms, sentiments, sentiment scores, etc.) as output. 

\begin{lstlisting}[caption={The subjects we used for Aspect-based Sentiment Analysis.}, captionpos=b,label={lst:keywords_ABSA},language=SQL, 
           showspaces=false,
           basicstyle=\footnotesize\ttfamily,
           frame=bt,
           commentstyle=\color{gray}]
zelensky, ukraine/ukrainian, russia/russian, putin, NATO
\end{lstlisting}

To process the raw results of the ABSA analysis, we use custom scripts in Python. In particular, for every tweet, we filter the aspect terms identified by {\em AWS Comprehend} to extract only the results, e.g., sentiments and the sentiment scores, related to our subjects (Listing~\ref{lst:keywords_ABSA}). In this way, for each subject (aspect term), we have all the tweets that mention it with the related sentiment.
In order to avoid multiple sentiments for one specific aspect in a single tweet, we define the following policies: 


\begin{enumerate}
  
  \item If there are 2 distinct sentiments for an aspect term:
  \begin{itemize}
  \item Neutral and Positive: the result is Positive
  \item Neutral and Negative: the result is Negative
  \item Negative and Positive: discard the sentiments
  \end{itemize}
  \item If there are 3 distinct sentiments for a specific aspect, we discard the sentiments.
\end{enumerate}

The  policies reported above ensure that each tweet  contributes  to exactly one sentiment for each aspect of our study. The result CSV files are sorted chronologically and plotted to show the aspects sentiments over time (e.g., hourly, daily, etc.). As the last post-processing step, we normalize the content (\#neutral, \#negative, \#positive) of the resulting CSV files to transform every value into a decimal between 0 and 1. This task facilitates comparing different graphs and extracting insights as we see in the next section in Figure~\ref{fig:ABSA_russia_vs_ukraine} and Figure~\ref{fig:ABSA_putin_vs_zelen}. It worth to mention that ABSA techniques still have shortcomings to be addressed by researchers. In our research, 10\% of the sentiments are missing because the {\em AWS Comprehend ABSA} could not detect the corresponding aspect terms from the input dataset.

\section{Results and Discussion}
\label{sec:results}
This section shows the results of both the quantitative and qualitative analyses we applied to our dataset to investigate the 2022 Russo-Ukrainian conflict as perceived  on Twitter and analyzed through the lenses of ABSA. In particular, we studied the volume of tweet messages related to the topic, both total and per user, to evaluate the impact of the conflict escalation on the Twitter community and to characterize the users involved. Then, we used Aspect-based Sentiment Analysis to investigate the users' sentiments about the considered subjects (reported in Listing~\ref{lst:keywords_ABSA}) and how they evolved over time. 

\subsection{Quantitative Analysis}
\label{sub:quantitative}
By executing the query described in Listing~\ref{lst:query}, we collected \tweets messages published by \users different users on Twitter during the considered period, i.e., one month before and one month after the triggering  of the armed conflict, that happened on February 24, 2022. 

To characterize the 2022 Russo-Ukrainian Conflict on Twitter, we analyzed our dataset using several metrics. First, we counted how many times our keywords were used in the collected tweets to identify the most popular subjects.  
\vspace{-0.5cm}
\begin{table}[]
\centering
\begin{tabular}{l|c|c|}
\cline{2-3}
& \textbf{\begin{tabular}[c]{@{}c@{}}Number of \\ Tweets\end{tabular}} & \textbf{\begin{tabular}[c]{@{}c@{}}Number of \\ Exclusive Tweets\end{tabular}} \\ \hline
\multicolumn{1}{|l|}{\textbf{Ukraine}}   & 1,907,016  & 1,202,640 \\ \hline
\multicolumn{1}{|l|}{\textbf{Putin}}     & 1,594,509  & 712,669 \\ \hline
\multicolumn{1}{|l|}{\textbf{Russia}}    & 1,364,853  & 512,946  \\ \hline
\multicolumn{1}{|l|}{\textbf{Russian}}   & 1,250,835  & 494,458  \\ \hline
\multicolumn{1}{|l|}{\textbf{Ukrainian}} & 544,308    & 166,628   \\ \hline
\multicolumn{1}{|l|}{\textbf{Zelensky}}  & 193,683    & 60,028  \\ \hline
\multicolumn{1}{|l|}{\textbf{NATO}}      & 330,638    & 0         \\ \hline
\end{tabular}
\caption{Number of Tweets returned by  our keywords}
\label{tab:tweets_per_keyword}
\vspace{-0.8cm}
\centering
\end{table}

We reported our results in Table~\ref{tab:tweets_per_keyword}, where the first column contains the overall number of tweets that include the highlighted keyword, while the second column contains the number of tweets where that keyword is used stand-alone, i.e., without the other six. \textit{Ukraine} is, by far, the most used keyword in our dataset, both in combination with others (34\% of the entire dataset), and alone (22\%). The discussion on Twitter also involved keywords related to Russia, that are \textit{Putin}, \textit{Russia}, and \textit{Russian}, respectively. Other keywords related to Ukraine are, instead, far less used. \textit{Zelensky}, for example, was only named in 3\% of the tweets collected (1\% alone). Overall, keywords related to Ukraine were used (alone) in 1,429,296 tweets, 26\% of the entire dataset, while keywords related to Russia were used in 1,720,073 tweets, 31\%. \textit{NATO}, instead, was named in 330,638 tweets, 6\% of the total number of tweets, always in combination with other keywords.
It is important to note that, by design, our dataset does not include any tweet with the \textit{NATO} keyword alone, as discussed in Section~\ref{sub:data_collection}. \\
\vspace{-0.3cm}
\begin{figure}
\centering
\includegraphics[width=0.9\columnwidth]{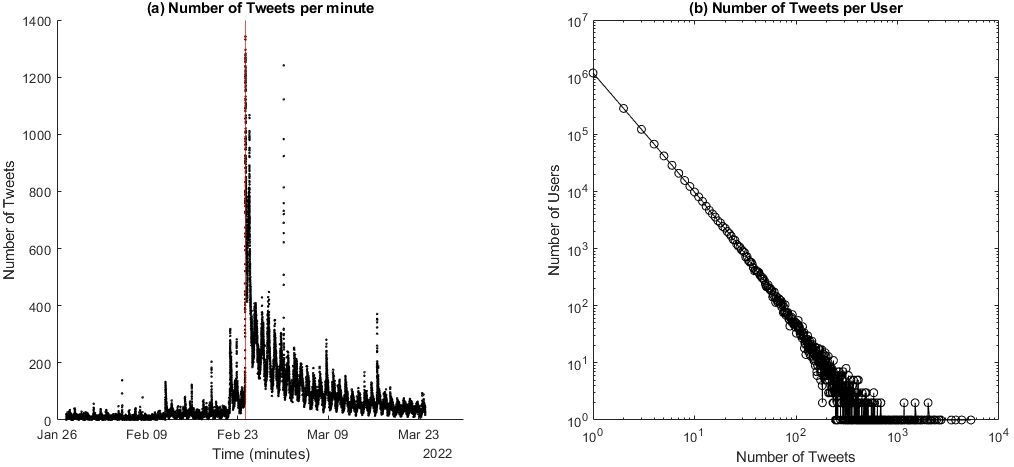}
\caption{Volume of tweets related to the 2022 Russo-Ukrainian Conflict, from January 27 to March 23. Overall number of tweets per minute (a), and number of tweets per account (b) in logarithmically scaled axes. The red vertical line in sub-figure (a) is February 24}
\label{fig:tweets_volume}
\end{figure}
\vspace{-0.3cm}
The volume of tweets related to the 2022 Russo-Ukrainian Conflict and published during the studied period is reported in Figure~\ref{fig:tweets_volume}. During the first month under consideration, the number of tweets per minute was steadily around 20, except for a few isolated spikes, as shown in Figure~\ref{fig:tweets_volume}.a. 
Then, it started to rise about a week before a central event in our dataset: the press conference at which the Russian president announced the start of a special military operation in Ukraine. On that date, the number of messages posted on Twitter suddenly shot up to over 1300 per minute. Subsequently, in the following weeks, it gradually decreased until it settled between 20 and 100 tweets per minute.

The vast majority of the users published very few tweets in the considered time window, as depicted in Figure~\ref{fig:tweets_volume}.b.  
This behaviour is perfectly in-line with what already noticed in the literature \cite{power_law_osn,emerg_prop}. After the user categorization discussed in Section~\ref{sub:pre_processing}, we looked at how much each category participated in the discussion about the escalation of the conflict. Table~\ref{tab:users_categorization} reports the number of users, the total number of tweets, and the average number of tweets per user for each category---and sub-category, if any---identified in our study. 
\begin{table}[]
\centering
\begin{tabular}{|lc|r|r|c|}
\hline
\multicolumn{2}{|c|}{\textbf{Account Category}} & \multicolumn{1}{l|}{\textbf{\begin{tabular}[c]{@{}l@{}}Number \\ of Users\end{tabular}}} & \multicolumn{1}{l|}{\textbf{\begin{tabular}[c]{@{}l@{}}Number \\ of Tweets\end{tabular}}} & \textbf{\begin{tabular}[c]{@{}c@{}}Average \\ Tweets per User\end{tabular}} \\ \hline
\multicolumn{2}{|l|}{\textbf{Baby  Accounts}}                            & 237,814   & 718,163   & 3.02  \\ \hline
\multicolumn{1}{|l|}{\multirow{3}{*}{\textbf{\begin{tabular}[c]{@{}l@{}}Trusted \\ Accounts\end{tabular}}}}  & \textbf{Celebrities}    & 50  & 2,075   & 41.50  \\ \cline{2-5} 
\multicolumn{1}{|l|}{} & \textbf{Very Popular}   & 690   & 11,461  & 16.61  \\ \cline{2-5} 
\multicolumn{1}{|l|}{}  & \textbf{Verified}  & 7,601  & 80,400  & 10.58 \\ \hline
\multicolumn{1}{|l|}{\multirow{3}{*}{\textbf{\begin{tabular}[c]{@{}l@{}}Abnormal \\ Accounts\end{tabular}}}} & \textbf{Zero Following} & 83,589   & 285,023  & 3.41  \\ \cline{2-5} 
\multicolumn{1}{|l|}{}   & \textbf{Fair FR}  & 28,323  & 55,821 & 1.97 \\ \cline{2-5} 
\multicolumn{1}{|l|}{}  & \textbf{Low FR} & 7,166  & 62,783 & 8.76  \\ \hline
\multicolumn{1}{|l|}{\multirow{2}{*}{\textbf{\begin{tabular}[c]{@{}l@{}}Unknown \\ Accounts\end{tabular}}}}  & \textbf{Suspended}      & 1,358 & 3,366   & 2.48 \\ \cline{2-5} 
\multicolumn{1}{|l|}{}  & \textbf{Deleted}   & 2,074  & 4,048  & 1.95 \\ \hline
\multicolumn{2}{|l|}{\textbf{Regular Accounts}}  & 1,489,940  & 4,360,028 & 2.93  \\ \hline
\end{tabular}
\caption{Number of users, tweets, and average tweets per user for each category}
\label{tab:users_categorization}
\vspace{-0.7cm}
\end{table}
The users who contributed most to the discussion are the trusted accounts. Specifically, celebrities tweeted on average 41 times, while very popular and verified accounts tweeted 16 and 10 times, respectively. This observation is not surprising, given that many news agencies that fall into this category. Instead, regular accounts, which are supposed to be owned and managed by normal users, tweeted on average only twice, in line with what can be seen in Figure~\ref{fig:tweets_volume}.b. It is important to mention how abnormal accounts, specifically those with low FR, differ significantly from regular accounts, having tweeted about eight times on average. These accounts could be involved in disinformation activities, and their analysis should be deepened in future work.
Another particularly interesting category is baby accounts. As described in Section~\ref{sub:pre_processing}, this category includes all the accounts created in the proximity of the considered time window, i.e., from January 27th to March 23rd. Therefore, it is very likely that a subset of those accounts was explicitly created for tweeting about the Russo-Ukrainian conflict.
\vspace{-0.5cm}
\begin{figure}
\centering
\includegraphics[width=0.8\columnwidth]{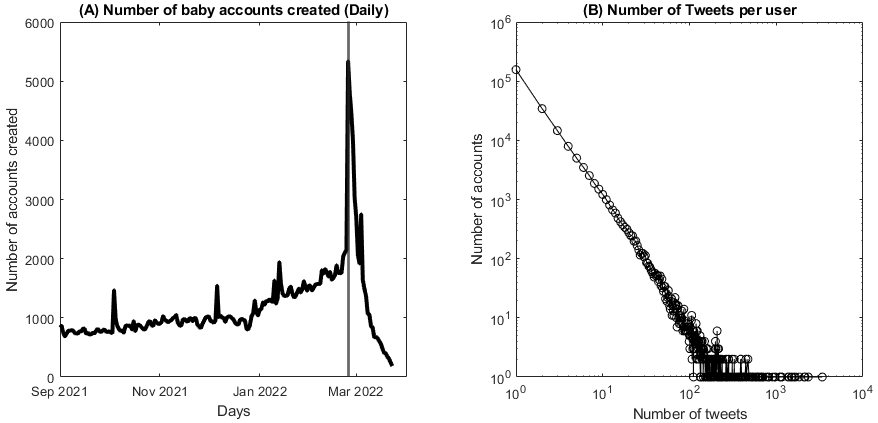}
\caption{Number of baby accounts created (Daily) from September 2021 to March 2022 (A), and number of tweets per baby account (B) in logarithmically scaled axes.}
\label{fig:baby_creation}
\vspace{-0.5cm}
\end{figure}

\noindent
Figure~\ref{fig:baby_creation}.A shows the number of accounts created over time. In the first period, around 1000 accounts were created per day. Then, the creation rate increases around 1 month before February 24---the grey vertical line in Fig.~\ref{fig:baby_creation}.A---to reach more than 5000 accounts created on that day. Then, the number suddenly decreases to values far less than the median before February 24.
Figure~\ref{fig:baby_creation}.B, instead, shows the number of tweets per baby account. Similarly to regular ones, the vast majority of baby accounts published very few tweets in the considered time window, following a power-law distribution.

\subsection{Qualitative Analysis}

Applying ABSA to the 2022 Russo-Ukrainian Conflict dataset (2 months tweets) provides an opportunity to extract and present hidden insights not easily achievable by other analytical techniques.
\vspace{-0.5cm}
\begin{figure}
\centering
\includegraphics[width=0.8\columnwidth]{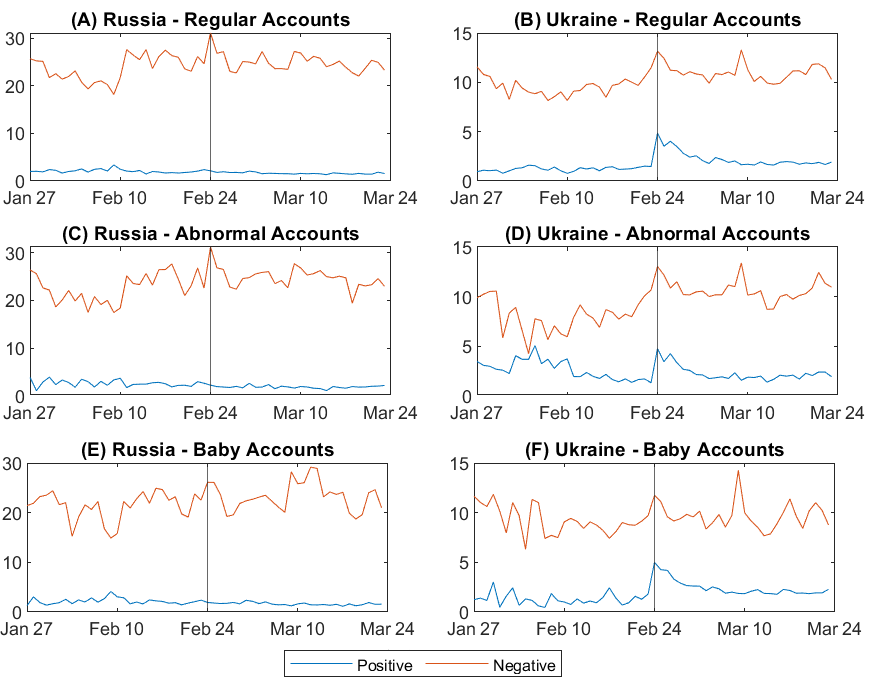}
\caption{Sentiment for the aspect terms 
``Russia'' and ``Ukraine'' over time. The grey vertical line is February 24.}
\label{fig:ABSA_russia_vs_ukraine}
\vspace{-0.5cm}
\end{figure}
The main objective is to investigate the sentiment of Twitter users about the main players involved in the escalation of the conflict and how that sentiment has changed over time. Given the space constraints, we only report the most relevant results, i.e.,  related to a subset of keywords and user categories---the complete outcomes of our analysis will be included in a future extension of this paper.
Figure~\ref{fig:ABSA_russia_vs_ukraine} illustrates our results for the ``Russia'' and ``Ukraine'' aspect terms over time, 1 month before and 1 month after the escalation of the conflict, i.e., February 24. Evolution of sentiments for the 2 considered aspects are quite different. For ``Ukraine'', the number of positive sentiments significantly increases after February 24. Instead, the same sentiment for ``Russia'' seems relatively constant over the two months considered, without experiencing any modification around February 24--this point will be expanded at the end of this section. On the other hand, the number of negative sentiments increase for ``Russia'' after the escalation of the conflict, while for Ukraine the negative sentiment increase rate is lower. 
For Abnormal Accounts, the graphs are significantly different. For the aspect term ``Ukraine'', positive sentiment rate is very high while negative sentiment rate is very low. Starting the conflict, negative sentiment rate increases.
\vspace{-0.5cm}
\begin{figure}
\centering
\includegraphics[width=0.8\columnwidth]{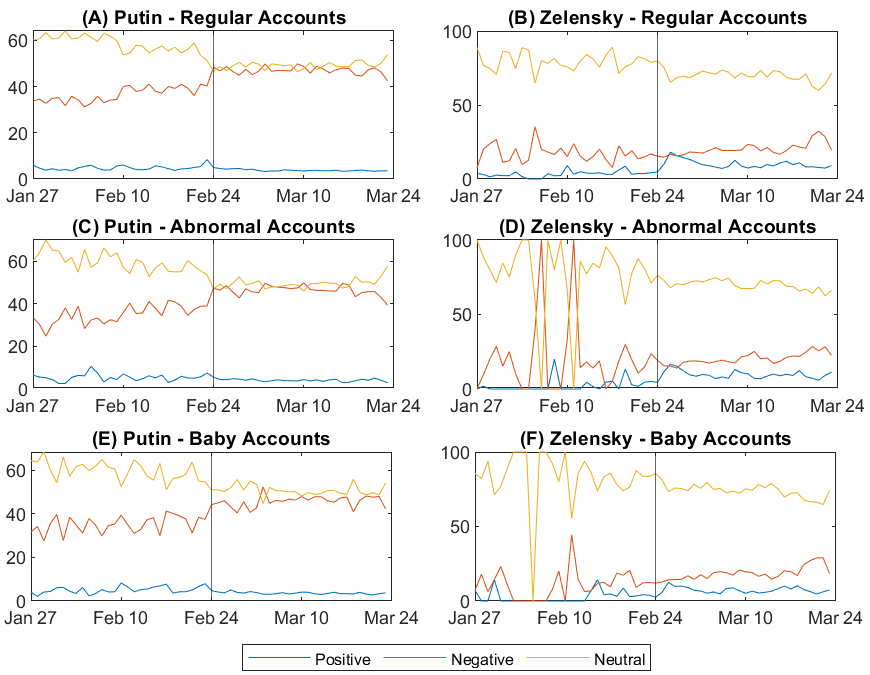}
\caption{Sentiment for the subjects ``Putin'' and ``Zelensky'' over time. The grey vertical line is February 24.}
\label{fig:ABSA_putin_vs_zelen}
\vspace{-0.5cm}
\end{figure}
Figure~\ref{fig:ABSA_putin_vs_zelen} illustrates the results for ``Putin'' and ``Zelensky'' aspect terms over time. The graphs and evolution patterns are completely different for the two aspect terms. For ``Putin'', neutral and negative rates converge after February 24. We believe that neutral sentiment mainly reflects news and information related to this conflict. After the conflict escalation, people feelings for ``Putin'' goes negative. Though, the positive rate is very low for both the aspect terms, and not only for the ``Putin'' one.

{\bf Findings.}
One of the main objectives of our qualitative analysis is discovering abnormal patterns within the chronological sentiments that may suggest disinformation activities, such as fake news and sentiments propagation, statistical manipulation, and, more in general, any disinformation campaign manipulating OSNs data, e.g., supporting a specific party, affecting public feelings, and possibly others.
In this regard, we can mention the sentiment trend for President Putin, shown in Figure~\ref{fig:ABSA_putin_vs_zelen}. For each category of accounts, it can be seen that the positive sentiment seems relatively constant throughout the period considered, with a very slight peak on 24 February. Instead, the Negative and Neutral trends appear to be counterbalanced. For almost the entire period considered, when one of the two rises, the other falls by approximately the same quantity, and vice versa. This would suggest that strong supporters of President Putin remained of the same opinion throughout the studied period, while neutral sentiments turned mainly negative following the escalation of the conflict. Though, it is very difficult to argument that there has been a massive campaign, by bot or paid collaborators,  or trolls in support of the Russian initiative, at it was stated in a few outlets\footnote{https://apnews.com/article/russia-ukraine-volodymyr-zelenskyy-kyiv-technology-misinformation-5e884b85f8dbb54d16f5f10d105fe850}. Another finding is the fact that President Zelensky, while having a constant exposure over standard media (TV, newspaper), has not been able to match the popularity on such media over Twitter. Indeed, while the general sentiment towards him is positive, it does not reach remarkable heights.
Finally, it worth noticing that, while the sentiment for President Putin has overall turned negative, the sentiment towards Russia has not suffered an equal decay in popularity. Maybe a sign that Twitter users are able to discriminate between the Government and the People the cited Government is supposed to serve.

{\bf Limitations.} This study includes some limitations that can be addressed in future works. First, since our analysis aims to identify disinformation in English-speaking countries, we considered only tweets in English. Analyzing data in other languages could enable the detection of disinformation targeting other countries, such as Russia and Ukraine.
Second, existing ABSA techniques are not able to discriminate different emotions of the same sentiment, e.g., madness and sadness, love and joy. As a results, if a specific tweet is, for example, negative for both Russia and Ukraine, our analysis does not reveal if the author is mad at Russia but sad for Ukraine, or vice versa. Using more advanced ABSA techniques will improve the effectiveness of this research. Finally, the user categorization is only based on a few user attributes, such as the number of following/followers. Using more advanced techniques could help to improve the existing categories and/or create new ones.  





\section{Conclusions and Future Work}
\label{sec:conclusion}
In this paper, we characterize the 2022 Russo-Ukrainian conflict by collecting and analyzing data from Twitter. Mainly, we investigated the data volume and the public perception of the conflict by leveraging statistical analysis techniques and Aspect-based Sentiment Analysis (ABSA). 
Our results do not support the hypotheses of a mass disinformation campaign, contrary to what claimed by a few mainstream media. However, we identified several anomalies in users' behavior and sentiment trends for some subjects that call for further research in the field. In particular, accounts with a low Friend Ratio (FR) twitted much more than regular users, with a sentiment trend for some keywords that diverged from other users. Also, even though baby accounts behave like regular users, their daily creation rate suggests that a subset of them was created specifically for tweeting about the conflict. For the cited reasons, baby and abnormal accounts might deserve a  thorough investigation using specialized bot detection techniques in future works.\\*
To the best of our knowledge, we are the first ones to use ABSA to analyze  the Twitter sentiment on the  Russo-Ukrainian conflict. While the findings and the used techniques are interesting on its own, this work also leave open a few relevant challenges. For instance,  whether ABSA is capable of a better expressivity with respect to standard mining and analysis techniques. As such, this work has the potential to pave the way for further research in the field.

\bibliographystyle{splncs04}
\bibliography{references}

\begin{thebibliography}{10}
\providecommand{\url}[1]{\texttt{#1}}
\providecommand{\urlprefix}{URL }
\providecommand{\doi}[1]{https://doi.org/#1}

\bibitem{bhatia2019comprehend}
Bhatia, P., Celikkaya, B., Khalilia, M., Senthivel, S.: Comprehend medical: a
  named entity recognition and relationship extraction web service. In: 2019
  18th IEEE International Conference On Machine Learning And Applications
  (ICMLA). pp. 1844--1851. IEEE (2019)

\bibitem{bhatia2020aws}
{Bhatia P. et al.}: Aws cord-19 search: A neural search engine for covid-19
  literature. arXiv preprint arXiv:2007.09186  (2020)

\bibitem{chen_ferrara}
Chen, E., Ferrara, E.: Tweets in time of conflict: {A} public dataset tracking
  the twitter discourse on the war between ukraine and russia. arXiv preprint
  arXiv.2203.07488  (2022)

\bibitem{chu2010tweeting}
Chu, Z., Gianvecchio, S., Wang, H., Jajodia, S.: Who is tweeting on twitter:
  human, bot, or cyborg? In: Proceedings of the 26th annual computer security
  applications conference. pp. 21--30 (2010)

\bibitem{emerg_prop}
Cresci, S., Di~Pietro, R., Petrocchi, M., Spognardi, A., Tesconi, M.: Emergent
  properties, models, and laws of behavioral similarities within groups of
  twitter users. Comput. Commun.  \textbf{150}(C),  47–61 (jan 2020),
  \url{https://doi.org/10.1016/j.comcom.2019.10.019}

\bibitem{ndiw2021}
Di~Pietro, R., Raponi, S., Caprolu, M., Cresci, S.: New Dimensions of
  Information Warfare, vol.~84. Springer International Publishing (2021).
  \doi{10.1007/978-3-030-60618-3}

\bibitem{gamzu2021identifying}
Gamzu, I., Gonen, H., Kutiel, G., Levy, R., Agichtein, E.: Identifying helpful
  sentences in product reviews. arXiv preprint arXiv:2104.09792  (2021)

\bibitem{10.1145/3110025.3110090}
Gilani, Z., Farahbakhsh, R., Tyson, G., Wang, L., Crowcroft, J.: Of bots and
  humans (on twitter). p. 349–354. ASONAM '17, Association for Computing
  Machinery, New York, NY, USA (2017),
  \url{https://doi.org/10.1145/3110025.3110090}

\bibitem{haq2022twitter}
Haq, E.U., Tyson, G., Lee, L.H., Braud, T., Hui, P.: Twitter dataset for 2022
  russo-ukrainian crisis. arXiv preprint arXiv:2203.02955  (2022)

\bibitem{kiriya2021troll}
Kiriya, I.: From “troll factories” to “littering the information
  space”: Control strategies over the russian internet. Media and
  Communication  \textbf{9}(4),  16--26 (2021)

\bibitem{nazir2020issues}
Nazir, A., Rao, Y., Wu, L., Sun, L.: Issues and challenges of aspect-based
  sentiment analysis: a comprehensive survey. IEEE Transactions on Affective
  Computing  (2020)

\bibitem{park2022voynaslov}
Park, C.Y., Mendelsohn, J., Field, A., Tsvetkov, Y.: Voynaslov: A data set of
  russian social media activity during the 2022 ukraine-russia war. arXiv
  preprint arXiv:2205.12382  (2022)

\bibitem{pohl2022twitter}
Pohl, J., Seiler, M.V., Assenmacher, D., Grimme, C.: A twitter streaming
  dataset collected before and after the onset of the war between russia and
  ukraine in 2022. Available at SSRN  (2022)

\bibitem{sivakumar2017aspect}
Sivakumar, M., Reddy, U.S.: Aspect based sentiment analysis of students opinion
  using machine learning techniques. In: 2017 International Conference on
  Inventive Computing and Informatics (ICICI). pp. 726--731. IEEE (2017)

\bibitem{xu2019bert}
Xu, H., Liu, B., Shu, L., Yu, P.S.: Bert post-training for review reading
  comprehension and aspect-based sentiment analysis. arXiv preprint
  arXiv:1904.02232  (2019)

\bibitem{power_law_osn}
Zang, C., Cui, P., Faloutsos, C., Zhu, W.: On power law growth of social
  networks. IEEE Transactions on Knowledge and Data Engineering
  \textbf{30}(9),  1727--1740 (2018). \doi{10.1109/TKDE.2018.2801844}

\end{thebibliography}

\end{document}